# Experimental and Numerical Investigation of Corona Discharge Induced Flow on a Flat Plate

Ravi Sankar Vaddi[1], Yifei Guan[1], Zhi yan Chen[1], Alexander Mamishev[2], Igor V Novosselov[1,3]

[1]Department of Mechanical Engineering, University of Washington, Seattle
[2]Department of Electrical and Computer Engineering, University of Washington, Seattle
[3]Institute for Nano-Engineering Systems, University of Washington, Seattle

*Abstract*— Electrohydrodynamic (EHD) flow induced by planar corona discharge in the wall boundary layer region is investigated experimentally and via a multiphysics computational model. The EHD phenomena has many potential engineering applications, its optimization requires a mechanistic understanding of the ion and flow transport. The corona EHD actuator consisting of two electrodes located in the wall boundary layer creates an EHD driven wall jet. The applied voltage between the electrodes is varied and the resulting effects in the charge density and flow field are measured. Constant current hotwire anemometry is used to measure velocity profile. The airflow near the wall acts a jet and it reaches a maximum of 1.7 m/s with an energy conversion efficiency of ~2%. The velocity decreases sharply in the normal direction. Multiphysics numerical model couples ion transport equation and the Navier Stokes equations to solve for the spatiotemporal distribution of electric field, charge density and flow field. The numerical results match experimental data shedding new insights into mass, charge and momentum transport phenomena. The EHD driven flow can be applied to flow control strategies and design of novel particle collectors.

## I. Introduction

Electrohydrodynamic flow also referred as corona wind or ionic wind has been studied in multiple applications such as convective cooling [1, 2], electrostatic precipitators [3-5], airflow control [6], surface particle trapping [7] and electroconvection [8, 9]. A high voltage corona electrode ionizes air molecules due to a strong electric field near the electrode. In a positive corona, negatively charged species return to the anode and position ions drift towards the cathode. The high kinetic energy ions transfer their momentum to air molecules outside the corona region through collisions initiating macroscopic air flow also known as ionic wind.

The scientific literature reports several corona configurations, such as wire to rod, point to ring, point to pot, wire to plate, coaxial cylinders for the generation of the corona. A quadratic relationship between the voltage and current characteristics and a linear relationship between the voltage and velocity are described in the literature, e.g., [10-12]. A surface corona discharge near the wall can be used to modify the airflow in the boundary layer, drag reduction, and separation control on airfoils at higher angles of attack. Velkoff et al. [13] have studied the effect of corona discharge on the laminar to turbulent transition point on a flat plate. The transition point shifted by 43 mm for an external free stream velocity of 53 m/s. Several researchers studied drag reduction using corona discharge on a flat plate [14, 15]. Moreau and Léger [6, 16] studied corona discharge on an inclined flat plate at low velocity and have observed a reduction in drag. A maximum velocity of 2.75 m/s for 500 µA current and the effect of an external free stream on boundary layer profile and pressure distribution has been studied experimentally [16]. However, studies on EHD flow development on a flat plate have been scarce.

To gain insight into the developing EHD boundary layer flow, Computational Fluid Dynamics (CFD) simulation that can couple flow, ion motion, and an electric field are required. The ion interaction with the air molecules can be modeled as an external force term in Navier Stokes equations (NSE). Most EHD models [17, 18] use an iterative approach to solve for the flow field and electric force. A modeling technique has proposed a new method to solve the modified NSE numerically [19]. A volumetric flux charge density is introduced in a finite volume; the conditions are determined from the experimental cathode current. The ionization boundary is defined by Peek's law [20], these are used as the thresholds for the onset of ionization

In this paper, we investigate the planar corona discharge in the wall boundary layer on a flat plate. The flow is studied experimentally and by the numerical simulations to resolve the spatiotemporal characteristics of ion concentration, velocity, and electric field. The electrical to kinetic energy transfer efficiency is calculated for both the model and the experiments.

## II. Experimental setup

The EHD flow is studied on a conducting flat plate with a shallow rectangular cavity. Fig. 1 shows the schematic representation of corona discharge induced flow in the wall boundary layer region on a conducting flat plate with a cavity. Its objective is to accelerate the flow near the wall to modify the boundary layer profile. The serrated edge copper electrode with a thickness of 0.2 mm serves as the anode. The pitch of the sawtooth is 5 mm. The ground electrode is a 1.5 mm thick steel rod. The flat plate is fabricated using 3D printing in polylactic acid polymer. A shallow rectangular cavity of 15 mm wide and 1.3 mm deep is built into the flat plate to aid ionization.

Corresponding author: Igor Novosselov
e-mail address: ivn@uw.edu

The anode overhangs the backward facing step of the cavity by 6 mm as shown in Fig. 2 such that the distance between the anode tips and cathode (L) is 9 mm. The width of the experimental setup is 100 mm. The cathode is flush mounted against the forward-facing step, see Fig. 1. The top of the electrode protrudes 0.2mm above the flat plate. The cathode protrusion results in the Stokes flow, not affecting the downstream velocity profile development.

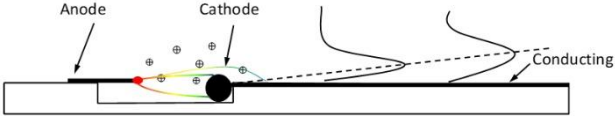

Fig. 1: Schematic diagram of the corona driven EHD wall jet; colored lines show electrical field line, solid down – velocity profile.

A variable voltage power supply (Bertan 205B-20R) is used to set the electrical potential between the electrodes. The cathode current is measured based on the voltage drop across a 1 MΩ resistor. The distance between the tip of the anemometer probe and the center of the cathode rod (D) is varied.

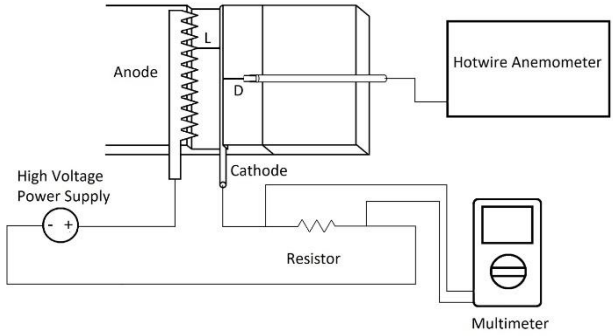

Fig. 2: Schematic of the experimental setup. A DC high voltage is applied between the copper sawtooth anode and a steel rod cathode, the distance between the anode and cathode is 9 mm

The anemometer was calibrated for a range of velocities 0.2 -5 m/s. The data is collected at a frequency of 20 kHz using a data acquisition module (myRIO-1900, National Instruments Inc.), the sampling time of 15s. The anemometer probe is mounted on an optical 3D stage; movement is controlled by micrometers with a range of 25 mm. The voltage between the anode and cathode is varied from 6 kV to 9 kV in 1 kV increments.

### III. MODELING

The interaction between the ion motion and the bulk flow is modeled by adding a body force to the momentum equation. The governing equations used in the model are:

$$\nabla \cdot \mathbf{u} = 0 \quad (1)$$

$$\rho \frac{D\mathbf{u}}{Dt} = -\nabla P + \mu \nabla^2 \mathbf{u} - \rho_e \nabla \varphi \quad (2)$$

where $\rho$, the air density (1.205 kg/m³) and $\mu$ is the air dynamic viscosity (1.846E-5 kg/(m-s)) are constant for incompressible isothermal flow, $\mathbf{u}$ is the velocity vector and $P$ is the static pressure. The equations for charge transport are:

$$\frac{\partial \rho_e}{\partial t} + \nabla \cdot \left[ (\mathbf{u} - \mu_b \nabla \varphi) \rho_e - D_e \nabla \rho_e \right] = S_e \quad (3)$$

$$\nabla^2 \varphi = -\frac{\rho_e}{\varepsilon_0} \quad (4)$$

where $\mu_b$ is the ion mobility (2.0E-4 m²/(Vs)) and $D_e$ is the ion diffusivity, $\varphi$ is the electric potential, $\rho_e$ is the charge density and $S_e$ is the volumetric force term calculated based on the experimental anode current. Instead of defining a thin surface within the computational domain to mark as the ionization zone boundary, a region with finite volume is determined by the electric field strength magnitude and constrained within 1mm of the anode tip. More details on the computational approach and treatment of each term can be found in the literature [19].

Commercial package ANSYS Fluent was used with custom subroutines model three-way coupling of ion motion, electric field, and fluid motion. Fig. 3 shows the schematic of the modeled geometry; the 2D assumption is used in the numerical simulation

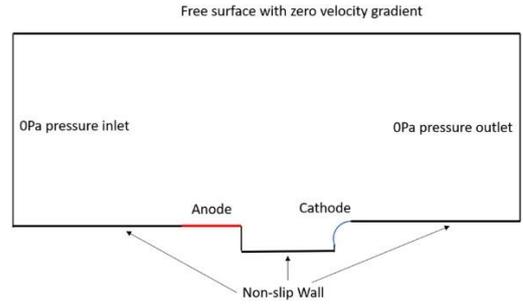

Fig. 3: Computational domain for the numerical simulation

### IV. RESULTS AND DISCUSSION

*A. Numerical Results*

The CFD models the process by which the ion-molecule collisions accelerate the bulk flow. Fig. 4 (a) shows the ion density contours. The ions are generated at the anode tip, and their motion is dominated by the electric field due to their high electric mobility, as the ion drift velocity is two orders of magnitude than the bulk flow [21]. The effect of space charge is observed as the charge density drifts upstream. The cathode recovers 85-90 % of the ion current that is generated, the other 10-15 % of charge species are recovered on the conducting plate reducing the parasitic effect of ions traveling upstream, as seen in the point to ring corona with insulating walls [19]. To parameterize the effect of the electrostatic force on the flow, the ratio of electric to the inertial force is defined as a non-dimensional parameter $X = \rho_e \varphi / \rho \mathbf{u}^2$ [19]. Fig. 4 (b) shows the velocity streamlines colored by the values of $X$, indicating the regions where the electric force is greater than the inertial force. The electric force is dominant between the cathode and anode where both the ion concentration and electric field strength are high. In the region downstream of cathode the effect of the ion interaction with the flow is minimal.

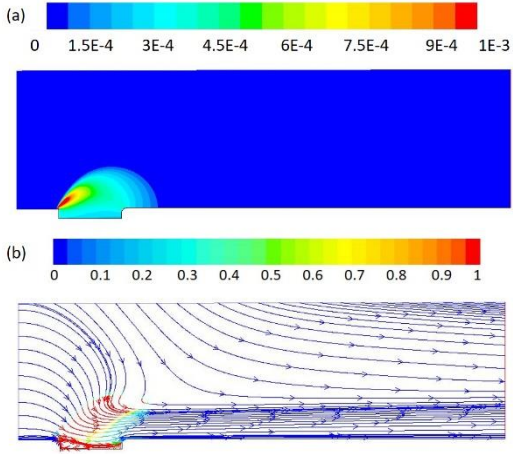

Fig. 4: Contour plots of the (a) charge density (C/m$^3$) and (b) streamlines by local $X$ (clipped to 1) for 8 kV case. Maximum - X=100

### B. Velocity Voltage Characteristics

The numerical results for $\varphi = 8$ kV case are compared with the experimental velocity profiles. The corona driven flow entrains gas in the vicinity, both streamwise and normal (impinging) components are present upstream of the cathode. Downstream of the cathode the normal component diminishes. The experimental and numerical velocity profile downstream of the cathode show similar trends exhibiting wall jet behavior [22-24]. The velocity reaches a maximum and then decays to near zero away from the plate. Both CFD and experimental data show the maximum velocity is 2mm from the surface. The maximum velocities in the numerical simulation is ~ 1.7 m/s and it is within the 10% of the experimental data.

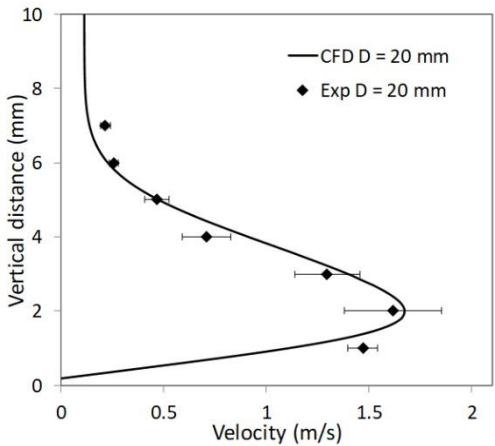

Fig. 5: Comparison of velocity profile between the experimental results and simulations at D = 20 mm of the EHD induced flow on a flat plate

### C. Energy Transfer Efficiency

The energy conversion efficiency can be calculated from the ratio of the kinetic energy flux to the electrical power produced by the corona discharge.

$$\eta = \frac{\frac{1}{2}\rho \int u^3 dA}{\varphi I} \quad (5)$$

The corona voltage and the anode current are obtained from the experiments. The kinetic energy flux is calculated for both experimental and numerical velocity profiles. TableTABLE I shows the values that are used for calculation. The energy transfer coefficient to the fluid is greater for external flow compared to internal flow due to entrainment of the surrounding fluid. Parasitic losses associated with the upstream motion of the ions and formation of the flow recirculation zone due to adverse pressure gradient are largely avoided by the introduction of conducting surface downstream of the cathode.

TABLE I: Comparison of electrical and kinetic power between the experiments and CFD

|  | 8 kV |
|---|---|
| I (μA) | 35 |
| $W_{K, Exp}$ (mW) | 6.1 ± 15.2% |
| $W_{K, CFD}$ (mW) | 5.9 |
| $W_E$ (mW) | 280 |

Energy transfer efficiency is 2.17 % ± 0.33% based on experimental and 2.1% based on CFD results. It was previously shown that the energy transfer efficiency is non-linear with respect to corona voltage [19]. The energy transfer efficiency for external flows is higher than in the point to ring internal flow (~ 1%) [19]. Further improvement in the energy transfer efficiency can be achieved by optimization of electrode geometry configuration, operational condition.

### IV. CONCLUSION

This work presents an experimental and numerical investigation of planar corona discharge in the wall boundary layer. The experimental data includes voltage, current, and velocity profile measurements. Multiphysics numerical simulation sheds insights into the interaction of force exerted by the motion of the ions in the electrical field on the airflow. The addition of charge flux as a generation of ions allows for the direct computation of electric body force. The numerical simulations agree with experimental data within 10%. The velocity profile of the corona driven is similar to a wall jet downstream of the cathode. Parametrization of the EHD wall jet and comparison with the traditional wall jet can be achieved using CFD modeling. The integrated velocity profile is used to calculate the electric to kinetic energy transfer efficiency. The efficiency is ~2%, which is greater than in the internal flow geometry due to higher flow entrainment. Energy transfer efficiency can be optimized by electrode geometry configuration, operational conditions.

ACKNOWLEDGMENT

This research was supported by the DHS Science and Technology Directorate, Homeland Security Advanced Research Projects Agency, Explosives Division and UK Home Office; grant no. HSHQDC-15-531 C-B0033, by the National Institutes of Health, grant NIBIB U01 EB021923.